\documentclass[preprint,twocolumn,preprintnumbers,aps,nofootinbib,tightenlines,floatfix,groupedaddress]{revtex4-1}

\usepackage{hyperref}
\usepackage{mathtools}
\usepackage{amsmath,amssymb}
\usepackage{cleveref}
\usepackage{graphicx}
\usepackage{xcolor}
\usepackage[utf8]{inputenc}
\usepackage{bm}
\usepackage{comment}
\usepackage[percent]{overpic}
\usepackage{qcircuit}
\usepackage{csquotes}
\usepackage[section]{placeins}
\usepackage{rotating}
\usepackage{csquotes}
\usepackage{scalerel}
\usepackage[export]{adjustbox}

\allowdisplaybreaks

\newcommand{\Tr}{\ensuremath{\text{Tr}}}

\newcommand\scalemath[2]{\scalebox{#1}{\mbox{\ensuremath{\displaystyle #2}}}}

\newcount\hour \newcount\hourminute \newcount\minute
\hour=\time \divide \hour by 60
\hourminute=\hour \multiply \hourminute by 60
\minute=\time \advance \minute by -\hourminute
\newcommand{\mydate}{\ \today \ - \number\hour :\number\minute}

\makeatletter
\renewcommand*\env@matrix[1][*\c@MaxMatrixCols c]{%
  \hskip -\arraycolsep
  \let\@ifnextchar\new@ifnextchar
  \array{#1}}
\makeatother

\begin{document}

\title{Fixed-Point Quantum Circuits for Quantum Field Theories}

\author{Natalie Klco}
\email{klcon@uw.edu}
\affiliation{Institute for Nuclear Theory, University of Washington, Seattle, WA 98195-1550, USA}
\author{Martin J.~Savage}
\email{mjs5@uw.edu}
\affiliation{Institute for Nuclear Theory, University of Washington, Seattle, WA 98195-1550, USA}

\date{\mydate}

\preprint{INT-PUB-20-003}

\begin{figure}[!t]
 \vspace{-1.5cm} \leftline{
 	\includegraphics[width=0.12\textwidth]{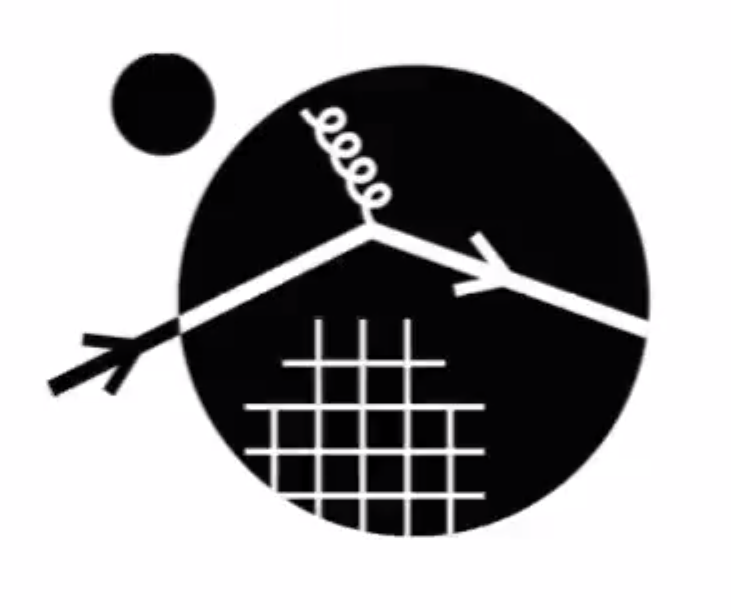}}
\end{figure}

\begin{abstract}
Renormalization group ideas and effective operators are used to efficiently determine
localized unitaries for preparing the ground states of non-interacting scalar field theories on digital quantum devices.
With these methods, classically computed ground states in a small spatial volume can be used to determine
operators for preparing the ground state in a beyond-classical
quantum register, even for interacting scalar field theories.
Due to the exponential decay of correlation functions and the double exponential suppression of digitization artifacts,  the derived quantum circuits are expected to be relevant already for near-term quantum devices.
\end{abstract}
\maketitle

\section{Introduction}
Quantum field theories (QFTs) describing the properties and dynamics of fundamental particles and
quantum many-body systems
 are anticipated to be addressed with
analog quantum simulation and digital quantum computation~\cite{Feynman:1981tf,Jordan:2011ne,Jordan:2011ci,Jordan:2014tma,Jordan:2017lea,Banuls:2019bmf}.
In addition to being essential to scientific applications in nuclear physics,
high-energy physics and basic energy sciences,
the distributed quantum degrees of freedom of QFTs provide a framework underlying the design of large-scale quantum simulators and fault tolerant quantum computers.
Numerical evaluations of QFTs typically involve
discretizing spacetime into a lattice of points
on which matter fields are defined,
with gauge fields defined on the links
between grid points.
Physical predictions can be derived from such calculations by extrapolating to the limit of infinite
spatial or spacetime volume and the continuum limit where the distance between grid points vanishes.
This can be accomplished by computing in sufficiently large volumes with sufficiently fine discretizations,
then extrapolating using known forms.

Preparing the ground state of a QFT Hamiltonian on digital quantum computers is challenging and has been identified as a leading contribution in estimates of the quantum resources required to simulate scattering in scalar field theory~\cite{2008arXiv0801.0342K,Jordan:2011ci,Jordan:2011ne,Jordan:2017lea}.
When spatially digitizing the scalar field, the efficiency of the quantum Fourier transform performed on each site can lead to a protection, through the Nyquist-Shannon sampling theorem, from induced systematic errors depending polynomially on the field digitization spacing~\cite{Jordan:2011ci,Jordan:2011ne,Somma:2016:QSO:3179430.3179434,PhysRevLett.121.110504,Macridin:2018oli,Klco:2018zqz}.

Preparing an arbitrary real function with support across the Hilbert space of a quantum register requires
an exponentially large number of entangling gates.
In the special case of a Gaussian profile, where the wave packet expands retaining its shape under time evolution, {\it Somma Inflation}
can be used to transform a Gaussian with support localized in the Hilbert space to a Gaussian with support distributed throughout the Hilbert space without an exponential increase in the number of entangling gates~\cite{Somma:2016:QSO:3179430.3179434}.
While individual Gaussians are a start, the scalar field ground state correlates Gaussians on each spatial site determined by the gradient operator.
Thus, introducing the spatial gradient operator creates entanglement among spatial sites,
significantly increasing the number of entangling gates that are required to prepare the ground state~\cite{Klco:2018zqz,PhysRevA.99.032306,Klco:2019xro,Klco:2019yrb}.
As previously shown, the number of required unitary operators for ground state preparation
scales linearly with the spatial volume
in massive QFTs due to the exponential localization of classical
correlations and entanglement~\cite{Klco:2019yrb}.
It was found that the rotation angles defining the unitary transformations become exponentially suppressed as the operators they define becomes increasingly non-local.

In this work, the properties of the ground states of non-interacting scalar field theories
and the symmetries of the corresponding quantum circuits are exploited
to derive spatially localized effective operators
for initializing large instances of ground-state wavefunctions, defining analytic \enquote{fixed points} of the localizable quantum circuits.
The systematic errors associated with these effective operators are shown to be exponentially suppressed.
In analogy with effective operators used in effective field theories induced or renormalized by removing degrees of freedom,
these effective operators are determined by integrating (tracing) over contributions from fields on lattice sites that are not involved in the controlled operations used to entangle the state on a given lattice site.
The unitary transformations associated with these effective operators rapidly evolve toward fixed points as the number of sites integrated out becomes large.
The implication is that relatively small systems with dimensions determined by the correlation length, and solved with classical computing resources, can be used
to determine the controlled unitary rotation operations required to prepare ground states
on a quantum computer capable of simulating a larger quantum field than classically possible.
This is also true for interacting scalar field theories.

\section{Scalar Field Ground State}
Working in the basis of eigenstates of the field operator $\hat{\phi}$~\cite{Jordan:2011ci,Jordan:2011ne},
the position-space, digitized wavefunction of a scalar field ground state mapped
onto a quantum register, $ |\Omega\rangle$ is,
\begin{align}
  |\Omega\rangle &= \sum_{\bm\phi} \psi(\bm\phi) |\bm\phi\rangle
  \ ,\
    \psi(\bm \phi) \propto e^{-\frac{1}{2} \bm\phi^T \bm K \bm\phi}
  \ ,
\end{align}
where the sum over $\bm\phi$ extends over the values of the field at each site for every spatial site.
This state
can be prepared from the fiducial state operationally with a non-unitary operator
$\hat{\Gamma}$
\begin{equation}
\hat \Gamma  =  e^{- {1\over 2} \hat {\bm\phi}^T {\bf K}  \hat {\bm \phi}} \ \ , \ \ |\Omega\rangle \propto \hat{\Gamma}H^{\otimes N} |0\rangle^{\otimes N} \ \ \ .
\label{eq:gamma}
\end{equation}
The wavefunction is sampled at regular intervals, $\delta_\phi$, between field truncations $\pm \phi_{\rm max}$ on each site of an $N$-site lattice with $n_Q$ qubits per site~\cite{Jordan:2011ne,Jordan:2011ci,Klco:2018zqz}.
Being constructed
from the field operators $\hat{\phi}_x|\phi_x\rangle = \phi_x |\phi_x\rangle$ at the site $x$,
$\hat{\Gamma}$ can be considered to be an operator with
wavefunction amplitudes along the diagonal in an infinite dimensional Hilbert space.
An alternate identification of $\hat{\Gamma}$ satisfying the preparation requirement of Eq.~\eqref{eq:gamma} is the density matrix of the digitized ground state.  Both perspectives will be useful in the identification of effective operators for the fixed-point circuit elements.

The symmetric matrix of site-wise correlations $\mathbf{K}$ is exponentially localized to the
diagonal with the form $e^{-mr}/r$, the two-site mutual information falls, up to power law components, as $e^{-2mr}$ with
separation, and the negativity is localized at the ultraviolet length scale of the finite-difference momentum
operator used to calculate $\mathbf{K}$~\cite{Srednicki:1993im,Audenaert:2002xfl,Klco:2019yrb}.
In explicit connection between the quantum circuitry necessary to prepare wavefunctions
and their intrinsic correlations, it has been shown that the circuit operations to prepare the
ground state of a massive scalar field can be localized~\cite{Klco:2019yrb}.
This localization is controlled by the structure of $\mathbf{K}$.
If the $\mathbf{K}$ matrix of correlations for a periodic lattice is truncated to be a band-diagonal matrix
with vanishing $K_{0i}$ for $i>d$, then quantum circuit elements can be made to depend only on the state of neighboring sites to a maximum distance $d$.
This feature of the circuit prevails in spite of the non-zero mutual information extending beyond this radius.
For clarity, in what follows the $\mathbf{K}$ truncation will be taken to be $d = 1$ unless otherwise noted and thus represent the largest contributions in the correlation hierarchy.

\section{Quantum Circuitry}
\begin{figure}[t]
  \centering
  \includegraphics[width=0.8\columnwidth,valign=c]{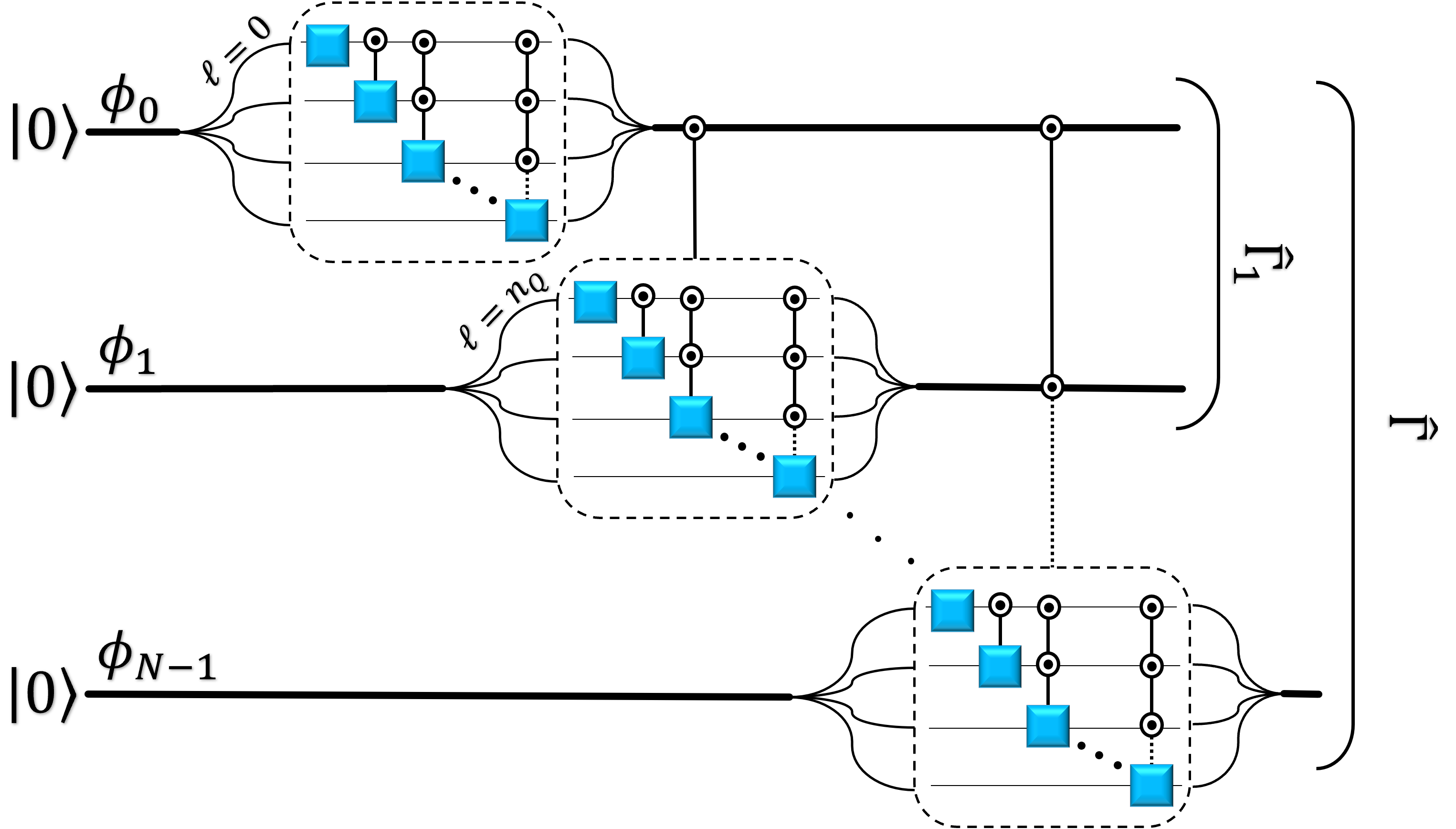}
  \includegraphics[width = 0.15\columnwidth,valign=c]{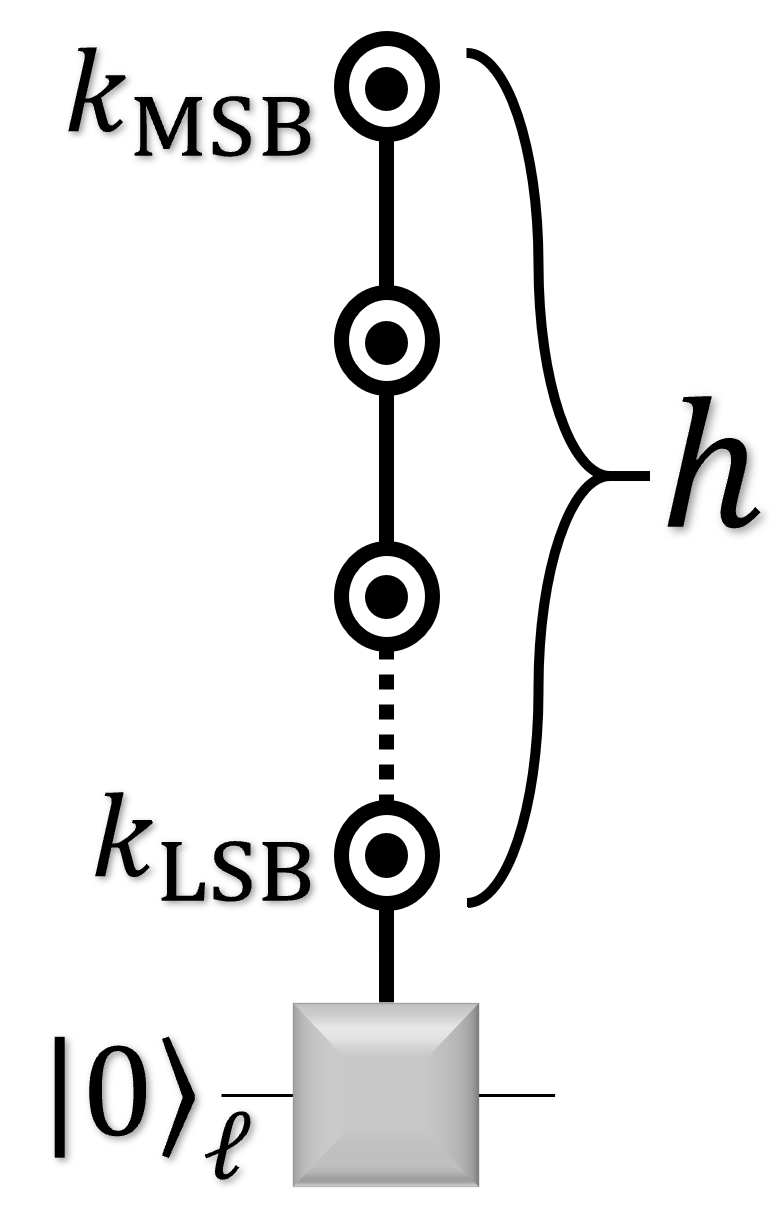}
  \caption{(color online) Quantum circuit for preparing an arbitrary real wavefunction.
  The open-closed controls indicate a set of operators controlled on each of the
  possible $2^h$ binary strings, $k$, with the most significant bit at the top.
  The blue squares indicate a corresponding set of $R_y$ rotations with unique angles $\theta_{\ell, k}$.}
  \label{fig:circuitDiagramThetas}
\end{figure}
\begin{figure}[t]
  \centering
  \includegraphics[width = 0.99\columnwidth]{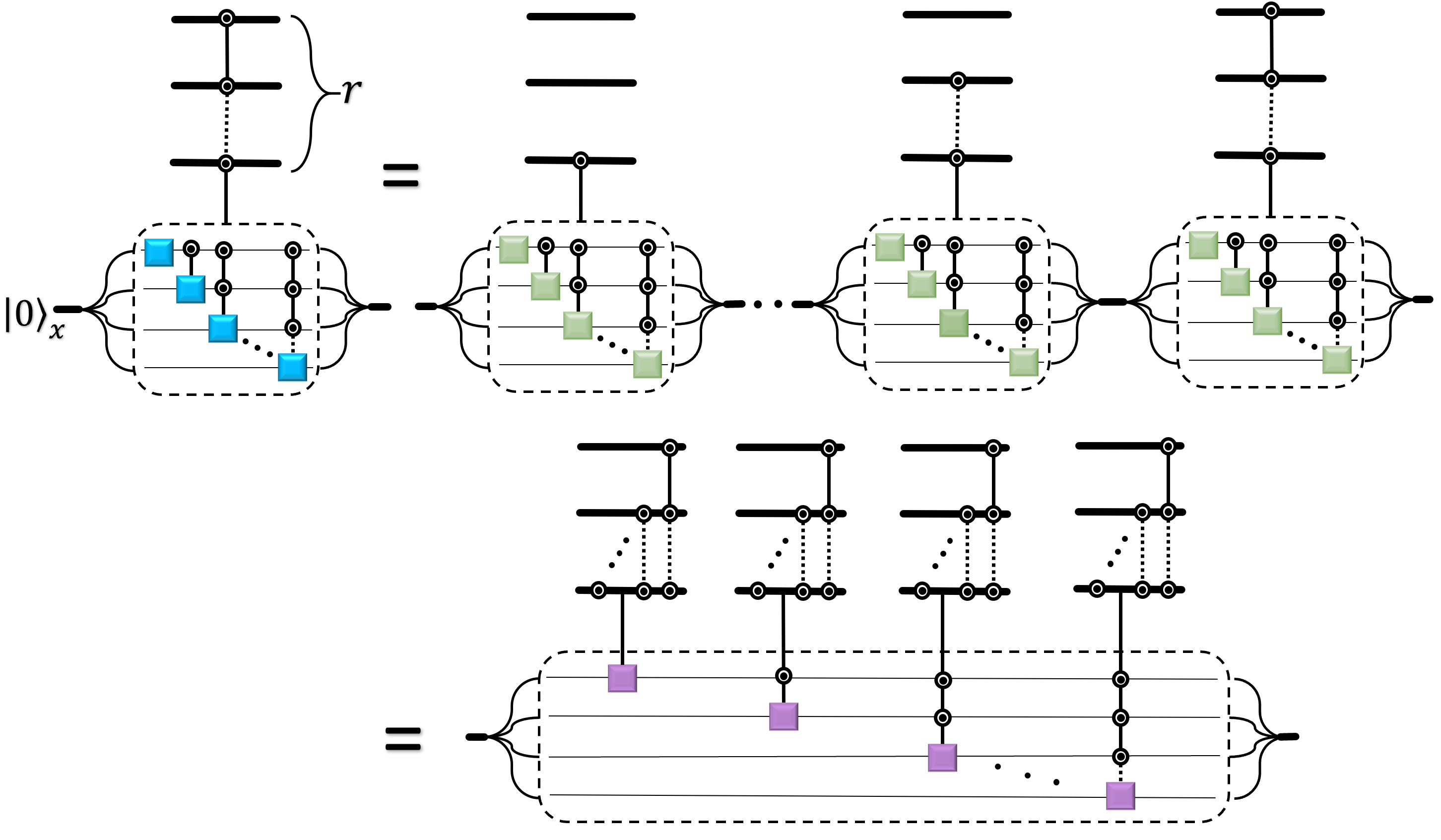}
  \caption{(color online) Site-wise expansion of angles $\theta_{\ell, k}$, appearing in rotation operators represented as blue squares, into angles $\alpha_{\ell h, k}$ appearing in rotation operators represented by purple squares.
Diagonal continuation dots represent the inclusion of controlled operators with
every intermediate number of controls.}
  \label{fig:sitewiseAlphaDiagram}
\end{figure}
By defining the structure of a quantum circuit, an implicit definition is made of the symmetries and wavefunction features that are naturally created upon implementation. Ideally, designed quantum circuits have structures that parallel those of the theory being simulated.  This allows, for example, decaying correlation functions to manifest as a hierarchy of operators with distance in the qubit register.  In the following, the circuit presented in Ref.~\cite{Klco:2019yrb} is first reviewed.  Having established physical properties of the field embedded in this structure, the discussion will then extend to determining fixed-point circuit parameters for ground state preparations larger than could be explicitly represented on classical devices.

Operators used in the localizable circuits can be defined by three numbers: two indices and an angle, as shown at the right of Fig.~\ref{fig:circuitDiagramThetas}.  The angle defines the rotation operator about the $y$-axis that acts on the target qubit.  The two indices $\ell$ and $k$, define the target qubit index and the binary-interpreted integer value of the controls.  Thus, declaring an angle of the form $\theta_{\ell, k}$ is equivalent to declaring a rotation operator along with its location and control values on previous qubits.
The $\theta$-circuit shown in Fig.~\ref{fig:circuitDiagramThetas} contains each of the $2^\ell$ controlled operators with differing values of $k$ acting on each qubit, ordered by $\ell$ from smallest to largest.
Operators with equal $\ell$ and differing $k$ commute.

Linear combinations of these $\theta$-angles produce $\alpha$-angles that are localized according to the two-point correlation function, and thus the exponentially decaying $\mathbf{K}$~\cite{Klco:2019yrb}.  These are denoted as $\alpha_{\ell h, k}^{x,r}$. The lower(upper) indices describe the associated controlled rotation at the level of qubits(sites).  There is a many-to-one relationship between the qubit- and site-indices, respectively.
For the qubit indices, $\ell$ and $k$ remain as defined for the $\theta$-angles.  The additional index $h$ is the \enquote{height} of the operator or number of controls.  For the $\theta$ circuit, all operators are defined to have $h = \ell$.  This equivalence between the operator height and its distance from the top of the qubit register is lifted for the $\alpha$ circuit in order to allow systematic localization of the operators and the relation becomes  $k_{\rm max} = 2^h-1$.
For the site indices, $x$ is the site number at which the rotation acts and $r$ is the site distance spanned by the operator.
These indices are connected by $x = \left\lfloor \ell/n_Q\right\rfloor$ and $r = \left\lfloor h/n_Q \right\rfloor$, where $\left\lfloor y \right\rfloor$ denotes the floor of $y$.

The site-wise $\alpha$-transformations are defined by expanding a $\theta$-operator extending over multiple sites defined by angles $\vec{\theta}^x_{\ell}$ into rotations over a truncated number of sites with
$h = \{ \ell - \left( x -1 \right) n_Q, \cdots, \ell - n_Q, \ell \}$
respectively being at physical distances $r = \{1, \ \cdots, x \}$.
This process is shown diagrammatically in Fig.~\ref{fig:sitewiseAlphaDiagram}.
At the left is a site-rotation controlled on the field at the $r$ previous sites.
It is desirable to isolate the sensitivity to controls at small spatial distance if the target wavefunction has localized correlations.
This is shown diagrammatically with the green circuit at the right of Fig.~\ref{fig:sitewiseAlphaDiagram} where site rotations controlled at short distances are extracted from those controlled at long distances.
A locality truncation removes circuit elements beginning at the right end of this circuit.
In the second line of this diagram, the purple circuit has grouped operators acting on a particular qubit.  Due to non-commutativity of rotations and controls, these angles are distinct from those of the previous expansion.
It is the angles of the latter, purple circuit $\alpha^{x,r}_{\ell h, k}$ that are naturally related to the $\theta_{\ell, k}$ on the left through simple linear combinations and averaging~\cite{Klco:2019yrb}.
The $\alpha$ angles for such decompositions with $\mathbf{K}$ truncated at $d=1$ are equivalent to the $\theta$-angles at distances 1, and vanishing for operators of greater spatial extent
\begin{equation}
  \alpha^{x,r}_{\ell h, k} = \begin{cases}
   \theta^x_{\ell,k} & r = 1\\
   0 & r > 1
  \end{cases} \ \ \ .
\end{equation}
This supports the statement that the $\alpha$-angles are controlled by the $\mathbf{K}$ matrix, exponentially suppressed in spatial locality controlled by the mass of the lightest particle.

The nature of these quantum circuits used to prepare the digitized ground state is such that adding
qubits to increase the density of states between the upper and lower values of the field at each site
changes the number and values of the angles associated with the unitary operations.
The additional field samples with each added qubit are interleaved with those at the previous field digitization, leading to a direct connection between the last qubit in each site and the high conjugate momentum modes.
With increasing $n_Q$, angles associated with the last qubit in each site tend towards a constant value of $\pi/4$, the angle for which additional samples are simply copies of their lower-digitization-scale predecessors.
This is a natural phenomena associated with the wavefunction being smooth or defined by an upper-bounded Fourier space.
In the circuit language, if $n_Q$ is taken to be large, the $\theta_{j n_Q-1, k}$-angles tend to $\pi/4$ for all integer $j \in \{1, ..., N\}$ and $k \in \{0, ..., 2^{j n_Q-1}-1\}$.
In the translation to localizable $\alpha$-angles shown in Fig.~\ref{fig:sitewiseAlphaDiagram}, these long-distance controlled operators tend towards a single-qubit operator $R_y(\pi/4)$ on the last qubit.
The $\alpha$-angles thus demonstrate a localization not only in spatial distance from exponentially-decaying spatial correlation functions, but also within the sites due to the hierarchy of conjugate momentum modes.
In contrast, operators acting on early qubits in the digitization (associated with the low conjugate momentum modes) yield rotation angles that tend to a fixed, non-constant distribution as a function of the field values on controlled sites above.
Given the expectation that small values of $n_Q\leq5$ will be sufficient for foreseeable calculations of the scalar field on quantum hardware, the following will be focused on fixed $n_Q$, where angles defining local operators evolve to fixed points as the number of spatial sites in the lattice becomes large.
Thus, for preparing lattices containing many correlation lengths, the number of unique local operators becomes independent of the lattice volume.

\section{Fixed Point Circuit}

For the particular case of the $\theta$-circuit (see Fig.~\ref{fig:circuitDiagramThetas}) where the state is prepared with asymmetric operations, the rotations on each site are dependent on the site register above the site on which the rotation acts.
The associated $\theta$-angles can be calculated from the ratio of sums of squared amplitudes in the wavefunction, marginalizing over the field on lower sites.  Continuing with the $\mathbf{K}$ truncation at $d = 1$, the $\theta$-angles are
\begin{equation}
  \theta_{\ell, k}^x = \arctan \sqrt{ \frac{\sum\limits_{j = 0}^{2^b-1} \langle \psi_{\rm n} | \hat{\Gamma}^{2,\rm{eff}}_{x} | \psi_{\rm n} \rangle }{\sum\limits_{j = 0}^{2^b-1} \langle \psi_{\rm d} | \hat{\Gamma}^{2,\rm{eff}}_{x} | \psi_{\rm d} \rangle}}\ \ \ .
  \label{eq:thetaangles}
\end{equation}
The states in the numerator and denominator are
\begin{equation}
\begin{aligned}
  |\psi_n\rangle &= \scalemath{0.95}{\left| \phi\left(\kappa^{\phi_{x-1}}\right), \phi\left( 2^{b+1}\kappa^{\phi_x}+2^b+j \right) \right\rangle} \\
  |\psi_d \rangle &=  \scalemath{0.95}{\left| \phi\left(\kappa^{\phi_{x-1}}\right) , \phi\left( 2^{b+1} \kappa^{\phi_x}+j \right) \right\rangle \ \ \ ,
}
\end{aligned}
\end{equation}
where $\phi(\kappa^{\phi_x}) \equiv \phi(x; \kappa^{\phi_x})$ denotes the field value at digitization address $\kappa^{\phi_x}$ at site $x$
\begin{equation}
\phi\left(\kappa^{\phi_x}\right) = \phi_{\rm max} - \delta_\phi \kappa^{\phi_x} \ \ \ .
\end{equation}
The site index and inner-site qubit index are
\begin{equation}
  x = \left\lfloor \frac{\ell}{n_Q} \right\rfloor \ , \ b = (n_Q-1)-\ell \bmod n_Q \ \ \ ,
\end{equation}
where the latter describes the location of the rotation within the site $x$ by the number of qubits back it acts from the last qubit in the site.
This is equal to the number of qubits within the site, but below the rotation, that need be additionally marginalized from the calculation.
The index $k$ has been interpreted in binary and grouped for its localization at each site of the scalar field
\begin{equation}
  k  = \{ \kappa^{\phi_0}, \kappa^{\phi_1}, \cdots, \kappa^{\phi_{x-1}} \}_2 \ \ \ .
\end{equation}
The site-wise $\kappa^{\phi_n}$ index takes values in the range $\{0, ..., 2^{n_Q}-1\}$~\footnote{As an example, for $n_Q = 2$, $x = 2$, and $k = 6$,
\begin{equation}
  k = 6 = 0110 = \{1,2\}_2 \ \ \ ,
\end{equation}
such that $\kappa^{\phi_0} = 1$ and $\kappa^{\phi_1} = 2$.  Thus, the control of $k = 6$ is associated with the scalar field on sites 0 and 1 of $\phi_0 = \phi_{\rm max}-\delta_\phi$ and $\phi_1 = \phi_{\rm max} - 2 \delta_\phi$.
}.
The effective operator, $\hat{\Gamma}^{2,\rm{eff}}_x$, relevant for calculating $\theta$ at site $x$ is a marginalization of $\hat{\Gamma}^2$ over the field values of all sites beneath the rotation location $x$,
\begin{equation}
\begin{aligned}
  \hat{\Gamma}_x^{2, \rm{eff}} &= \Tr_{\bm\phi_{>}}  \left[ \hat{\Gamma}^2 \right] \\  \bm\phi_> &= \{\phi_{x+1}, \phi_{x+2}, \cdots, \phi_{N-1}\} \ \ \ .
\end{aligned}
\end{equation}
Note that the reduced $\hat{\Gamma}^{2,{\rm eff}}_x$ operator marginalizes in the probabilities, not the amplitudes.    This is a manifestation of the use of $y$-axis rotations in the circuit construction, producing trigonometric functions of the rotation angles in the wavefunction amplitudes.  Sums over squares of these amplitudes then sequentially remove qubits from the end of the register.  In terms of the qubit reduced density matrix, the angles are defined by ratios of the diagonal matrix elements
\begin{equation}
 \theta_{\ell, k} = \arctan \sqrt{\frac{\langle 2k+1 | \hat \rho_{\ell} | 2 k +1 \rangle }{ \langle 2k | \hat{\rho}_{\ell} |2 k \rangle  }}\ \ ,
\end{equation}
with $\hat{\rho}_\ell$ the ground state reduced density matrix of the first $\ell+1$ qubits
\begin{equation}
  \hat{\rho}_\ell = \Tr_{q > \ell} \left[ \hat{\rho} \right] \ \ \ .
\end{equation}
By marginalizing at the qubit level (rather than at the site level as done in Eq.~\eqref{eq:thetaangles}), the sums in the numerator and denominator over the $2^b$ values of the field at higher digitization frequency in the site $x$ are handled in the trace reduction of the density matrix.
For the current application of initializing the ground state, the operator definition in Eq.~\eqref{eq:gamma} is more efficient for the construction of local effective operators.

If the qubit registers in the lower region of the circuit are taken to be continuous fields without field truncation,
\begin{equation}
\begin{aligned}
  \hat{\Gamma}_x^{2,{\rm eff}} &\rightarrow \int_{\bm\phi_{>}} \hat{\Gamma}^2 \\
  &\propto e^{- {\hat{\bm\phi}_{\leq}}^T \mathbf{K}_{x+1} {\hat{\bm\phi}_{\leq}} + K_{01}^2\frac{ \det \mathbf{K}_{\bar{x}-1} }{ \det \mathbf{K}_{\bar{x}}} \hat{\phi}_{x}^2 } \ \ \ ,
\end{aligned}
\label{eq:effectiveGammaContinuous}
\end{equation}
where $\mathbf{K}_{n}$ is the first $n \times n$ sub-block in $\mathbf{K}$ and $\bar{x} = N-(x+1)$ is the number of sites below site $x$ (the number of sites in $\phi_>$).
Note that for the calculation of $\theta$-angles with $\mathbf{K}$ truncation at $d=1$, the field values above $\hat{\phi}_{x-1}$ may be ignored as they cancel in the ratio in Eq.~\eqref{eq:thetaangles}.  In this scenario, the effective operator reduces to a local operator
\begin{multline}
  \hat{\Gamma}^{2,\rm{eff}}_x \propto e^{- K_{00} \hat{\phi}_{x-1}^2 - 2 K_{01} \hat{\phi}_{x-1} \hat{\phi}_x} \ \times \\* e^{ - \left( K_{00} - K_{01}^2\frac{ \det \mathbf{K}_{\bar{x}-1} }{ \det \mathbf{K}_{\bar{x}}}\right) \hat{\phi}_{x}^2  }
   \ .
  \label{eq:effectiveGammaContinuousLocal}
\end{multline}
This operator captures the effective $\mathbf{K}$ matrix of correlations relevant for the unitaries acting at site $x$, integrating out degrees of freedom in a manner resembling the renormalization group.
With $d>1$, the effective operator remains exponentially localized with the structure of $\mathbf{K}$.
The ratio of determinants tends to a constant at large lattice sizes ($N \rightarrow \infty$), and can be expressed for continuous $\bar{x}$ fields as,
\begin{align}
  \frac{ \det \mathbf{K}_{\bar{x}-1} }{ \det \mathbf{K}_{\bar{x}}
  } &
  = \scalemath{0.72}{\frac{2}{K_{00} + \eta \left( 1 + \frac{2}{-1+(K_{00}-\eta)^{-\bar{x}} (K_{00}+\eta)^{\bar{x}}} \right)} } , \label{eq:detratioXdependent}
   \\*
  & \hspace{-0.3cm} \xrightarrow[\bar{x} \to \infty]{} \frac{2}{K_{00} \left( 1 + \sqrt{1 - \frac{4 K_{01}^2}{K_{00}^2}} \right) }\ ,
  \label{eq:detratioLimit}
\end{align}
with $\eta = \sqrt{K_{00}^2 - 4K_{01}^2}$.
At large volumes, these two elements of the correlation matrix are
\begin{align}
  K_{00}^\infty &= \frac{2  \sqrt{4 + m^2}}{\pi} \ E\left( \frac{4}{4+m^2} \right) \ \ \ , \label{eq:K00infinity} \\
  K_{01}^\infty &= \frac{\sqrt{4 + m^2}}{3 \pi} \left[m^2 K\left(\frac{4}{4+m^2} \right)  \right. \label{eq:K01infinity} \\* & \hspace{1.6cm} \left.-(2 + m^2) E\left( \frac{4}{4 + m^2}\right)  \right] \ , \nonumber
\end{align}
where $K$ and $E$ are the complete elliptic integrals of the first and second kind, respectively.
Having analytic forms of the asymptotic $\mathbf{K}$ matrix elements and
determinant ratio is a convenient but unnecessary feature of the non-interacting scalar field. They are computationally inexpensive and can be easily calculated for large volumes.

\begin{figure*}
  \includegraphics[width=0.48\textwidth]{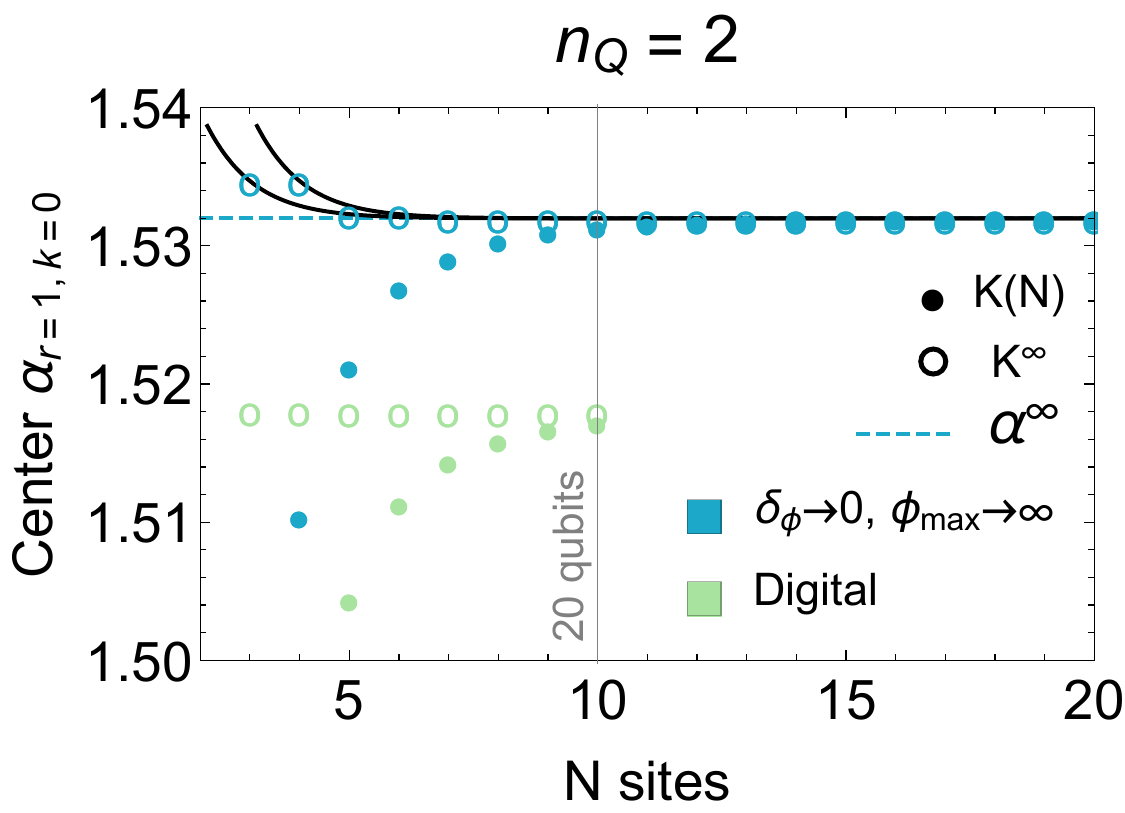}
  \includegraphics[width=0.48\textwidth]{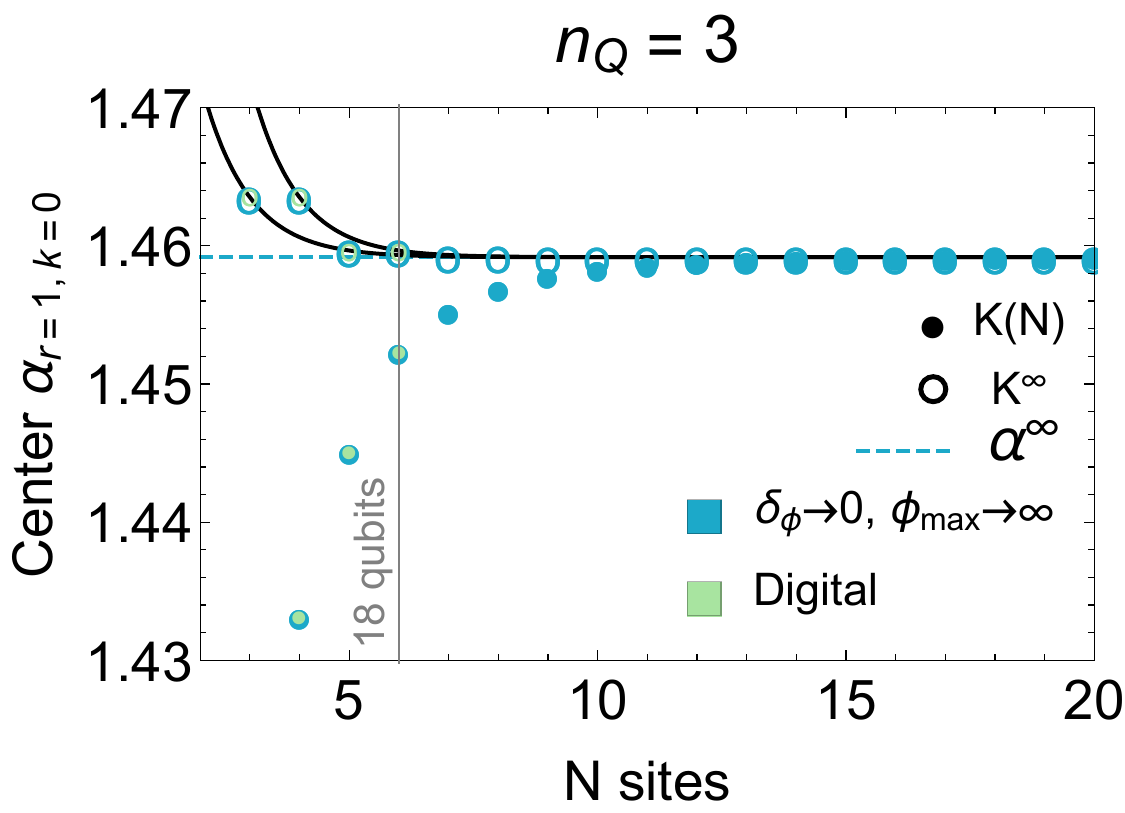}
  \caption{Evolution of an angle associated with a localized operator at the center of the lattice, $\alpha\left( x = \left\lfloor \frac{N}{2} \right\rfloor, r = 1, \ell = n_Q\left( \left\lfloor \frac{N}{2} \right\rfloor + 1 \right) -1, h = 2 n_Q-1, k = 0\right) $, with lattice size $N$. The scalar field is defined by $m = 0.3$, truncated in correlations at $K_{01}$, and truncated for digitization with maximum field value $\phi_{\max} = 3.5$.  Blue points are the results of calculations approximating sites in the second half of the lattice to be continuous fields without field truncation. Green points are the results of calculations within a digitized and truncated qubit representation.  Closed circles indicate the $\mathbf{K}$ matrix scales with system size, while open circles indicate that the infinite-volume $\mathbf{K}$ matrix elements are used. The solid black lines show the $N$ dependence of the continuum calculations with infinite-volume $\mathbf{K}$ matrix elements, while the blue dashed line is the fixed-point $\alpha^\infty$ calculated for continuum sites in the lower half lattice in the limit of infinite volume.}
  \label{fig:centeralphaconvergence}
\end{figure*}
The determination of fixed-point circuit elements has occurred in the limit of infinite volume, infinite field truncation $\phi_{\rm max}$, and continuous quantum registers on marginalized lattice sites of the field.
Leading corrections to the above expressions due to the finite extent of the lattice scale as
${\cal O}\left( e^{-m N}\right)$ up to polynomial factors scaling approximately as $1/\sqrt{N}$.
To quantify the systematic uncertainties associated with assumed continuous quantum registers, consider the distribution upon marginalization of the field at a single lattice site,
\begin{equation}
  \mathcal{B}_{1}(\phi_\ell) = \sum\limits_{\phi_c} e^{- \left( K_{00} \phi_c^2 + 2 K_{01} \phi_\ell \phi_c \right)} \ \ \ .
\end{equation}
Utilizing Poisson resummation to relate the $\phi$-symmetrized Dirac comb, producing digitized field samples, to a sum over its Fourier modes, deviations are found to be exponentially suppressed,
\begin{align}
  &\mathcal{B}_1^\infty(\phi_\ell) = \sqrt{\frac{\pi}{K_{00} \delta_\phi^2}} e^{\frac{K_{01}^2 \phi_\ell^2}{K_{00}}} \times \\* &
  \scalemath{0.95}{\left[ 1 + 2 \sum_{n > 0 } e^{-\frac{n^2 \pi^2}{K_{00} \delta_\phi^2}} \cos \left( n \pi \left(1 + \frac{2 K_{01} \phi_\ell}{K_{00} \delta_\phi} \right)\right) \right] \ ,} \nonumber
\end{align}
where the term in brackets can written as an elliptic theta function.
For fields digitized onto qubits with $\delta_\phi = \frac{2 \phi_{\rm max}}{2^{n_Q} -1}$, the deviations from the continuum scale as $\mathcal{O}\left(e^{-2^{2n_Q}} \right)$, double exponentially in the number of qubits.

This rapid convergence is another manifestation of the Nyquist-Shannon sampling theorem, the effects of which can be seen in Fig.~\ref{fig:centeralphaconvergence}, where the convergence of a rotation angle in the center of the lattice is shown.
The green points have been calculated through representation of the $2^{N n_Q}$ dimensional digitized wavefunction for systems of up to 20 qubits.
The blue points have had the effective operator replaced by the continuum and untruncated (in $\phi_{\rm max}$) effective operator of Eq.~\eqref{eq:effectiveGammaContinuousLocal}.
As such, the $N$ dependence of the blue closed points comes from the determinant ratio in Eq.~\eqref{eq:detratioXdependent} and the $N$ dependence of the $\mathbf{K}$ matrix itself.
If the infinite volume values of $\mathbf{K}$, as shown in Eqs.~\eqref{eq:K00infinity} and~\eqref{eq:K01infinity}, are used, the open points and black lines are recovered, demonstrating a rapid convergence to the continuum angles.  The continuum values, $\alpha^\infty$, are shown as blue dashed lines which are calculated by defining the effective operator with Eqs.~\eqref{eq:detratioLimit},~\eqref{eq:K00infinity}, and~\eqref{eq:K01infinity}.

On the left panel of Fig.~\ref{fig:centeralphaconvergence}, a coarse qubit digitization of $n_Q = 2$ is used on each site.  The angle calculated in the continuum without field truncation agrees with that calculated in the digitized space to $\sim 1\%$.
Thus, substituting the fixed-point $\alpha$-angles for the digitized circuit provides sufficiently precise determinations of rotation angles necessary for initializing the ground state on even small instances of near-term quantum devices (where this precision matches that expected on hardware).
Due to the double exponential convergence in the number of qubits used to digitize the field, increasing $n_Q$ to 3 qubits (right panel of Fig.~\ref{fig:centeralphaconvergence}) shows good agreement between the angles defining the continuum and digitized circuits.
Differences of angles in small volumes, where wavefunctions can be represented classically, are found to be $\sim0.001\%$.
While increasing $n_Q$ requires additional circuit operations to prepare the ground state (see Fig.~\ref{fig:sitewiseAlphaDiagram}), the number of gates, $\mathcal{O}\left( N 2^{2n_Q}\right)$, grows more slowly than the ability to improve them.
The double exponential convergence in digitization artifacts implies that increasing $n_Q$, and thus the fidelity of the wavefunction, parametrically improves the fidelity-to-gate ratio when using fixed-point circuits to initialize the scalar field ground state.

As the distance truncation, $d$, of $\mathbf{K}$ is raised, the effective operator relevant for calculating the $\theta$-angles at a particular site becomes less local.  The modification to the reduced $\mathbf{K}$ matrix in the effective operator is generally
\begin{equation}
  \hat\Gamma^{2,\rm{eff}} \propto e^{-\phi_{\leq}^T \left( A - B C^{-1} B^T\right) \phi_{\leq}}\ \ ,
\end{equation}
with
\begin{equation}
\scalemath{0.9}{
  \mathbf{K} = \begin{pmatrix}[c|c]
    A & B  \\
    \hline
    B^T & C  \\
  \end{pmatrix} \ , \  \left[ A \right] = (x+1) \times (x+1) \ .
}
\end{equation}
It can be seen that the correction to the site-site correlations in the effective operator, $B C^{-1} B^T$, is non-zero only in the lower subblock of dimension $d$ controlled by the largest non-zero $K_{0d}$ retained in $\mathbf{K}$. This is consistent with the physical intuition that the off-diagonal elements of $\mathbf{K}$ control the site distance of communication throughout the lattice.
This localization to the lower sub-block is also connected to the localization of the effective operator(s) necessary to calculate $\theta$-angles defining the fixed-point circuit.

\section{Reflections}

In this work, fixed-point quantum circuits have been introduced for the preparation of the non-interacting scalar field ground state on digital quantum hardware.
Determining the circuit elements necessary to initialize large instances of the quantum field requires classical computational resources scaling only with the spatial distance of correlations.
For the massive scalar field, these correlations decay exponentially with distance, leading to the ability to determine fixed point circuits for preparing the ground state on quantum devices for larger lattices than could be stored classically.  This technique is also applicable to interacting scalar field theory.

In this work, the continuum limit (decreasing lattice spacing) has not been considered. Taking this limit is required to make predictions for physical observables with a complete quantification of uncertainties.  As the lattice spacing is reduced, the number of lattice sites within a correlation volume increases.  This scaling is power law with the lattice spacing. The fixed-point analysis that we have presented remains valid, but with an increased number of required $\alpha$-angles.

While it is conceived that the ground state of an interacting theory can be initialized beginning from the non-interacting ground state adiabatically, there is no barrier to applying these fixed point methods to interacting ground states as well.
This provides an alternative state preparation mechanism that avoids additional circuit depth scaling with unpredictable spectral gaps throughout the dynamical adiabatic process.
While such applications evade analytic solution, both perturbative corrections to the circuit elements and non-perturbative analyses can be performed.
The perturbative approach leverages the analytic control demonstrated here in defining the non-interacting state preparation circuit, though
further exploration is necessary to understand the corrections when interactions are strong.
Alternatively, it is viable to inform fixed point circuits for interacting ground states non-perturbatively. Infinite volume limits of circuit elements can be reliably extrapolated from finite volume calculations capturing only the exponentially localized correlation length scale.

Fixed-point quantum circuits are expected to be relevant for initializing the ground states of fields defined by locally-interacting massive particles with exponentially decaying correlation functions or area-law entanglement.  It is further anticipated that confining gauge theories will admit fixed-point quantum circuits, suggesting how classical calculations of the QCD vacuum could inform state preparation on beyond-classical quantum devices.

\begin{acknowledgments}
We would like to thank Aidan Murran for inspiring interactions.
NK and MJS were supported by the Institute for Nuclear Theory with DOE grant No. DE-FG02-00ER41132, and Fermi National Accelerator Laboratory
PO No. 652197.   This work is supported in part by the U.S. Department of Energy, Office of Science, Office of Advanced Scientific Computing Research (ASCR) quantum algorithm teams program, under field work proposal number ERKJ333.
NK was supported in part by a Microsoft Research PhD Fellowship.
\end{acknowledgments}

\bibliography{fpcbib}
\end{document}